\documentclass[10pt]{article}
\def \ccomma{\raise 2pt\hbox{,}} % Le petit livre de TeX page 234
\def \D {\hbox{d}}
\def \Pn     {{\rm P_{\rm n}}}
\def \PVI    {{\rm P_{\rm VI}}}
\def \PV     {{\rm P_{\rm V}}}
\def \PIV    {{\rm P_{\rm IV}}}
\def \PIII   {{\rm P_{\rm III}}}
\def \PII    {{\rm P_{\rm II}}}
\def \PI     {{\rm P_{\rm I}}}
\def\AnnENS{Ann.~\'Ec.~Norm.~}

\def \tg   {\mathop{\rm tg}\nolimits}
\def \cotg {\mathop{\rm cotg}\nolimits}

\def \bfF {{\bf F}}
\def \bfN {{\bf N}}
\begin{document}

\title{On a surface isolated by Gambier}

\author{Runliang Lin${}^{1}$ and Robert Conte${}^{2,3}$
{}\\
\\ 1. 
   Department of mathematical sciences, Tsinghua University, 
\\ Beijing 100084, P.R. China.
\\	
   2. Centre de math\'ematiques et de leurs applications,
\\ \'Ecole normale sup\'erieure de Cachan, CNRS, 
\\ Universit\'e Paris-Saclay, 
\\ 61, avenue du Pr\'esident Wilson, F--94235 Cachan, France.
\\
\\ 3. Department of Mathematics, The University of Hong Kong,
\\ Pokfulam Road, Hong Kong.
\\
\\    E-mail RLin@mail.tsinghua.edu.cn, Robert.Conte@cea.fr
}

\maketitle

\begin{abstract}
We provide a Lax pair for the surfaces of Voss and Guichard,
and 
we show that such particular surfaces considered by Gambier
are characterized by a third Painlev\'e function.
\end{abstract}

\noindent \textit{Keywords}:
Surfaces of Voss and Guichard;
Lax pair;
Painlev\'e III.

\noindent \textit{2000 Mathematics Subject Classification:} 
   33E17, % Painlev\'e-type functions
   34A05, % ODEs. Explicit solutions and reductions
   34Mxx. % Differential equations in the complex domain [See also 30Dxx, 32G34] 

% http://www.ams.org/msc/msc2010.html
% http://www.ams.org/msc/classification.pdf
%\noindent \textit{AMS MSC 2000} % http://www.ams.org/msc/classification.pdf
%30D30, % Meromorphic functions, general theory
%33E17. % Painlev\'e-type functions
%33E05, % Elliptic functions and integrals
%33E30, % Other functions coming from differential, difference and integral equations
%34A05, % ODEs. Explicit solutions and reductions
%35A20, % PDEs. Analytic methods, singularities
%35C05, % PDEs. Solutions in closed form
%35Q58, % Other completely integrable equations
%35Q55, % PDEs. NLS-like equations
%35Q99  % Equations of mathematical physics, none of the above
%37J30  % Dynamical systems. Obstructions to integrability
%37J35  % Finite-dim Hamiltonians. Completely integrable, integration methods

%\noindent \textit{PACS 2001} % http://publish.aps.org/PACS/01pacs.html
                             % http://publish.aps.org/PACS/pacs_01.asc
% 02.30.H  %  Differential equations, ordinary
% 02.30.Jr,%  Differential equations, partial
% 02.30.Ik,%  Integrable systems,
% 02.30.Ks,%  Delay and functional equations
% Where are the discrete equations?
% 05. Statistical physics, thermodynamics, and nonlinear dynamical systems
% 05.45.-a Nonlinear dynamics and nonlinear dynamical systems
% 42.65    % Nonlinear optics
% 42.65.J  % Nonlinear optics, self-focusing
% 42.65.Wi.% Nonlinear wave guides
% 45 Classical mechanics of discrete systems

%\noindent \textit{PACS}
% (cosmology, general relativity, nonlinear ODEs, quantum gravity)
%02.30.+g   % Mathematical methods in physics
           %   Function theory, analysis

\vfill\eject
\tableofcontents
%\vfill\eject
% ============================================================================
\section{Introduction. Surfaces of Voss and Guichard}
% ============================================================================

\bigskip

Let us first recall two equivalent definitions of these surfaces:
a geometric one and an analytic one.
Our notation follows the review of Gambier \cite{Gambier1931}.

Geometrically, the surfaces of Voss \cite{AVoss1888} and Guichard \cite{Guichard1890} 
are by definition those which admit a conjugate net made of geodesics.
For instance, every minimal surface is such a surface.

Analytically,
they can be characterized by their three fundamental quadratic forms
$ \D \bfF.\D \bfF$,
$-\D \bfF.\D \bfN$,
$ \D \bfN.\D \bfN$,
in which $\bfF(u,v)$ is the current point of the surface
and $\bfN(u,v)$ a unit vector normal to the tangent plane.
Choosing the coordinates $(u,v)$ defined by the geodesic conjugate net,
these are \cite[p.~362]{Gambier1931}
\begin{eqnarray}
& &
\left\lbrace
\begin{array}{ll}
\displaystyle{
{\rm I}=X_u^2 \D u^2 + 2 \cos(2 \omega) X_u Y_v \D u \D v + Y_v^2 \D v^2,\
}\\ \displaystyle{
{\rm II}=\sin(2 \omega) (X_u \D u^2 + Y_v \D v^2),\
}\\ \displaystyle{
{\rm III}=\D u^2 + 2 \cos(2 \omega) \D u \D v + \D v^2,\
}
\end{array}
\right.
\label{eqFundamental_forms}
\end{eqnarray}
with the notation $X_u=\partial X(u,v)/\partial u$, \dots,
they depend on three functions $\omega, X, Y$ of two variables,
and $2 \omega$ is 
the angle between the two conjugate geodesics.
It is remarkable that,
among the three Gauss-Codazzi equations \cite[p.~362]{Gambier1931}
\begin{eqnarray}
& &
\left\lbrace
\begin{array}{ll}
\displaystyle{
\omega_{uv}-\frac{1}{2}\sin(2 \omega)=0,
}\\ \displaystyle{
X_v - \cos(2 \omega) Y_v=0,
}\\ \displaystyle{
Y_u - \cos(2 \omega) X_u=0,
}
\end{array}
\right.
\label{eqGC}
\end{eqnarray}
the first one characterizes the surfaces with a constant total (Gauss) curvature. 

% ============================================================================
\section{Their Lax pair}
% ============================================================================

Gambier succeeded in introducing a deformation parameter $\lambda$,
thus upgrading the moving frame equations to a Lax pair,
but he did not write this Lax pair explicitly, so let us do it here.

The moving frame equations (Gauss-Weingarten equations)
only depend on the coefficients of the first and second fundamental forms,
and the spectral parameter is introduced, 
as in the case of surfaces with a constant mean curvature,
 by noticing the
invariance of the Gauss-Codazzi equations (\ref{eqGC})
under the scaling transformation $(u,v) \to (\lambda u, \lambda^{-1} v)$. 
The traceless Lax pair is
\begin{eqnarray}
& & {\hskip -15.0 truemm}
\partial_u \psi= L \psi,\
\partial_v \psi= M \psi,\
\\ & & {\hskip -15.0 truemm}
%\matU=\begin{pmatrix} a11 & a12 \cr a21 & a22 \cr \end{pmatrix} NEW (amsmath)
%\matU=       pmatrix{ a11 & a12 \cr a21 & a22 \cr }             OLD
L= \pmatrix{
\displaystyle{ \frac{2 X_{uu}}{3 X_u}+\frac{2 X_u}{3 X_v} \cotg(2 \omega) \omega_v }
                                                  & 0 & Y_u \tg(2 \omega) \cr
\displaystyle{           -     \frac{2 Y_v}{\lambda^2 Y_u} \cotg(2 \omega) \omega_u} &
\displaystyle{ -\frac{  X_{uu}}{3 X_u}-\frac{4 X_u}{3 X_v} \cotg(2 \omega) \omega_v} 
 & 0 \cr
\displaystyle{           -         \frac{1}{\lambda^2 Y_u} \cotg(2 \omega)}  &
\displaystyle{                               \frac{1}{Y_v} \cotg(2 \omega)}  & 
\displaystyle{ -\frac{  X_{uu}}{3 X_u}+\frac{2 X_u}{3 X_v} \cotg(2 \omega) \omega_v} 
 \cr
	},
\\ & & {\hskip -15.0 truemm}
M= \pmatrix{
\displaystyle -\frac{  Y_{vv}}{3 Y_v}-\frac{4 Y_v}{3 Y_u} \cotg(2 \omega) \omega_u 
&
\displaystyle           -     \frac{2 \lambda^2 X_u}{X_v} \cotg(2 \omega) \omega_v & 0 \cr
0 & 
\displaystyle  \frac{2 Y_{vv}}{3 Y_v}+\frac{2 Y_v}{3 Y_u} \cotg(2 \omega) \omega_u 
&
\displaystyle                                    X_v        \tg(2 \omega) \cr
\displaystyle                             \frac{1  }{X_u} \cotg(2 \omega)  &
\displaystyle            -        \frac{\lambda^2  }{X_v} \cotg(2 \omega)  & 
\displaystyle -\frac{  Y_{vv}}{3 Y_v}+\frac{2 Y_v}{3 Y_u} \cotg(2 \omega) \omega_u 
 \cr
	},
\end{eqnarray}
with the zero-curvature condition, 
\begin{eqnarray}
& & {\hskip -15.0 truemm}
\left\lbrack \partial_u - L,\partial_v-M \right\rbrack 
= \pmatrix{       -F E_1         & E E_1          & -(E G-F^2) E_2 \cr 
                  -G E_1         & F E_1          & -(E G-F^2) E_3 \cr 
									 G E_2 - F E_3 & -F E_2 + E E_3 & 0\cr 
  }=0,
\end{eqnarray}
denoting $E_j,j=1,2,3$ the lhs of (\ref{eqGC}),
and $E,F,G$ the coefficients of the first fundamental form,
\begin{eqnarray}
& &
E=X_u^2,\ F=X_u Y_v \cos(2 \omega),\ G=Y_v^2.
\end{eqnarray}

% ============================================================================
\section{Surfaces applicable on a surface of revolution}
% ============================================================================

Gambier \cite[p.~99]{Gambier1928} investigated surfaces
whose first fundamental form I, Eq.~(\ref{eqFundamental_forms}),
has coefficients 
$X_u$, $Y_v$, $\omega$ only depending on the 
single variable $x=u+v$.
Denoting for shortness $X_u=\xi$, $Y_v=\eta$,
he first obtains
\begin{eqnarray}
& &
\D X=\xi \D u +(\xi+2 C_1) \D v,\
\D Y=(\eta + 2 C_2) \D u +\eta \D v,\
\nonumber\\ & &
\xi  + 2 C_1 = \eta \cos(2 \omega),\
\eta + 2 C_2 = \xi  \cos(2 \omega),\
\label{eqGambier1928Eq5p99} 
\end{eqnarray}
with $C_1$, $C_2$ two integration constants. 
After a possible conformal transformation, this defines two reductions
of the Gauss-Codazzi equations,
either
\begin{eqnarray}
& &
\frac{\D^2 \omega}{\D x^2}= \frac{m^2}{2} \sin (2\omega),\
m=\hbox{arbitrary constant},\
\label{eqGambier1928EqAp100}  
\end{eqnarray} 
or
\begin{eqnarray}
& &
\frac{\D^2 \omega}{\D x^2}= \frac{e^x}{2} \sin (2\omega).
\label{eqGambier1928EqBp100}  
\end{eqnarray} 

The first reduction (\ref{eqGambier1928EqAp100})
integrates with elliptic functions and
is handled in full detail by Gambier \cite[pp.~100--105]{Gambier1928}.
\medskip

As to the second reduction (\ref{eqGambier1928EqBp100}),
Gambier unexpectedly fails to integrate it.
This ordinary differential equation (ODE) 
belongs to the class of second order first degree ODEs
\begin{eqnarray}
& &
\frac{\D^2 u}{\D x^2}+ \sum_{j=0}^{3} A_j(x,u) \left(\frac{\D u}{\D x}\right)^j=0,
\label{eqClassOrder2Degree1GroupPoint}
\end{eqnarray}
whose property is to be form-invariant under 
the group of point transformations
\begin{eqnarray}
& & {\hskip -15.0 truemm}
(u,x) \to (U,X):\
u=\varphi(X,U),\ x=\psi(X,U),\ U=\Phi(x,u),\ X=\Psi(x,u). 
\label{eqGroupPoint}
\end{eqnarray}
Roger Liouville \cite{RLiouville} enumerated equivalence classes of 
(\ref{eqClassOrder2Degree1GroupPoint})
\textit{modulo} the group (\ref{eqGroupPoint})
but,
as later pointed out by Babich and Bordag \cite{BB1999},
he forgot the important class,
to which the ODE (\ref{eqGambier1928EqBp100}) belongs,
when the invariants which he denotes $\nu_5$ and $w_1$ both vanish.

When $\nu_5$ and $w_1$ both vanish,
the coefficients $A_3$, $A_2$, $A_1$ in
the class (\ref{eqClassOrder2Degree1GroupPoint})
can be set to zero by a transformation (\ref{eqGroupPoint}),
thus defining
the five remarkable four-parameter nonautonomous ODEs
% =================================================================
\begin{eqnarray*}
%   \PVI\  : \
& & 
\frac{\D^2 U}{\D X^2}=\frac{(2 \omega)^{3}}{\pi^2}
\sum_{j=\infty,0,1,x}\theta_j^2 \wp'(2\omega U+\omega_j,g_2,g_3),
\\ % \PV\   : \ 
& & 
\frac{\D^2 U}{\D X^2}=
-2 \alpha \frac{\cosh U}{\sinh^3 U}
-2 \beta \frac{\sinh U}{\cosh^3 U}
-2 \gamma e^{2 X} \sinh (2 U)
-\frac{1}{2} \delta
 e^{4 X} \sinh (4 U),
\\ %\PIII\ : \ 
& & \frac{\D^2 U}{\D X^2}=
 \frac{1}{2} e^   X (\alpha e^{2 U} + \beta  e^{-2 U})
+\frac{1}{2} e^{2 X}(\gamma e^{4 U} + \delta e^{-4 U}),
\\ %\PIV'\ : \ 
& & \frac{\D^2 U}{\D X^2}=
-\alpha U +\frac{\beta}{2 U^3}+\gamma\left(\frac{3}{4}U^5+ 2 X U^3 + X^2 U \right)
+2 \delta (U^3+X U),
\\ %\PII'\ : \ 
& & \frac{\D^2 U}{\D X^2}=\delta (2 U^3 + X U) + \gamma (6 U^2 + X) + \beta U+ \alpha,
\end{eqnarray*}
% =================================================================
in which the summation in the first equation runs over the four half-periods $\omega_j$
of the Weierstrass elliptic function $\wp$.

The third one is precisely, up to rescaling,
the ODE (\ref{eqGambier1928EqBp100}) isolated by Gambier,
and the main result of Ref.~\cite{BB1999} is the existence 
of a point transformation
mapping these five four-parameter ODEs to 
the representation of the Painlev\'e equations
chosen by Garnier
\cite{GarnierThese} \cite{CMBook}
(i.e.~five equations with four parameters,
the last one unifying $\PII$ and $\PI$),
\begin{eqnarray}
\PVI\ : \ u'' &=&
\frac{1}{2} \left[\frac{1}{u} + \frac{1}{u-1} + \frac{1}{u-x} \right] u'^2
- \left[\frac{1}{x} + \frac{1}{x-1} + \frac{1}{u-x} \right] u'
\nonumber \\ & &
+ \frac{u (u-1) (u-x)}{x^2 (x-1)^2}
  \left[\alpha + \beta \frac{x}{u^2} + \gamma \frac{x-1}{(u-1)^2}
        + \delta \frac{x (x-1)}{(u-x)^2} \right],
\nonumber \\ \PV\ : \ u'' &=&
\left[\frac{1}{2 u} + \frac{1}{u-1} \right] u'^2
- \frac{u'}{x}
+ \frac{(u-1)^2}{x^2} \left[ \alpha u + \frac{\beta}{u} \right]
+ \gamma \frac{u}{x}
+ \delta \frac{u(u+1)}{u-1},
\nonumber\\
\PIII\ : \ u'' &=&
\frac{u'^2}{u} - \frac{u'}{x} + \frac{\alpha u^2 + \gamma u^3}{4 x^2}
 + \frac{\beta}{4 x}
 + \frac{\delta}{4 u},
\label{P6P0def}
\label{eqPIII}
\\ \PIV'\ : \ u'' &=&
\frac{u'^2}{2 u} + \gamma \left(\frac{3}{2} u^3 + 4 x u^2 + 2 x^2 u\right)
+ 4 \delta (u^2 + x u) - 2 \alpha u + \frac{\beta}{u},
\nonumber \\ \PII'\ : \ u'' &=&
\delta (2 u^3 + x u) + \gamma (6 u^2 + x) + \beta u + \alpha.
\nonumber 
\end{eqnarray}
The point transformations which realize this mapping are, respectively, 
\begin{eqnarray*}
%   \PVI\  : \ 
& & x      =\frac{              e_3-e_1}{e_2-e_1}\ccomma\
u=\frac{\wp(2 \omega U,g_2,g_3)-e_1}{e_2-e_1},\
\\ %\PV\   : \ 
& & x=e^{2 X},\ u=\coth^2 U,\ % t=\coth^2 T,\ 
\\ %\PIII\ : \ 
& & x=e^{2 X}  ,\ u=e^X e^{2 U}     ,\ % t=e^X e^{2 T},
\\ %\PIV'\ : \ 
& & x=X      ,\ u=U^2      ,\ % t=T^2,
\\ %\PII'\ : \ 
& & x=X      ,\ u=U        .% \ t=T,
\end{eqnarray*}

Therefore the mapping between the ODE (\ref{eqGambier1928EqBp100}) for $\omega(x)$
and the third Painlev\'e equation (\ref{eqPIII}) for $u(\xi)$ is
either
\begin{eqnarray}
& &
e^{2 i \omega}=2 \alpha e^{-x} u,\ \xi=-\frac{1}{4 \alpha \beta}e^{2 x},\
\alpha \beta \not=0,\ \gamma=0,\ \delta=0, 
\label{eqChoice1}
\end{eqnarray} 
or equivalently
\begin{eqnarray}
& &
e^{2 i \omega}=\gamma e^{-x} u^2,\ \xi=\sqrt{-\frac{1}{\gamma \delta}} e^{x},\ 
\alpha=0,\ \beta=0,\ \gamma \delta\not=0.
\label{eqChoice2}
\end{eqnarray} 

As is well known, the third Painlev\'e equation  
has three kinds of solutions:

(i) two-parameter transcendental solutions, which is the generic case, 
and one cannot proceed beyond the description of Gambier \cite[pp.~105--106]{Gambier1928};

(ii) one-parameter Riccati-type solutions,
but for our case $\gamma \delta\not=0$ this does not happen;

(iii) zero-parameter rational\footnote{Algebraic solutions of (\ref{eqPIII}) \cite{MurataP3} 
are in fact rational solutions for another representative of $\PIII$
in its equivalence class under $(u,x) \to (g(x) u,f(x))$, with $f(x)=\sqrt{x}$, $g(x)=1$.
All algebraic solutions of $\Pn$ equations, $n=2,3,4,5$,
are similarly rational.} solutions,
the only ones being,
with the choice (\ref{eqChoice1}),
\begin{eqnarray}
& &
u=\left(-\frac{\beta}{\alpha} \right)^{1/2} \xi^{1/2},\ \gamma=0,\ \delta=0,
\label{eqRationalChoice1}
\end{eqnarray} 
or equivalently with the choice (\ref{eqChoice2}),
\begin{eqnarray}
& &
u=\left(-\frac{\delta}{\gamma} \right)^{1/4} \xi^{1/2},\ \alpha=0,\ \beta=0. 
\label{eqRationalChoice2}
\end{eqnarray} 

However, these rational solutions correspond to $\sin(2 \omega)=0$,
forbidden because the second fundamental form would vanish. 
Consequently, all solutions of (\ref{eqGambier1928EqBp100}) are transcendental.

% ============================================================================
\section{Future developments}
% ============================================================================

The equation (\ref{eqGC})${}_1$ for constant total curvature surfaces
(sine-Gordon equation)
possesses many closed form solutions which obey neither 
(\ref{eqGambier1928EqAp100}) nor
(\ref{eqGambier1928EqBp100}),
for instance the factorized solution \cite{Steuerwald}
\begin{eqnarray}
& &
\tg \frac{\omega}{2}= \frac{J_1(u+v)}{J_2(u-v)}\ccomma
\end{eqnarray} 
in which $J_1$ and $J_2$ are Jacobi elliptic functions,
a degeneracy of which is
\begin{eqnarray}
& &
\tg \frac{\omega}{2}= \frac{\sin k(u+v)}{\sin k(u-v)}\ccomma
\end{eqnarray} 
or the $N$-soliton solution \cite{SG-N-soliton},
which depends on $2 N$ arbitrary constants.
The difficulty to build Voss-Guichard surfaces from such solutions
is the integration of the linear system (\ref{eqGC})${}_{2,3}$ for $X(u,v)$
and $Y(u,v)$.

Another useful development would be to find a Darboux transformation for the system
(\ref{eqGC}).

% =======================================================================
\section*{Acknowledgments}

This work was partially funded by the National Natural Science Foundation 
of China grant 11471182, and the Hong Kong GRF grant HKU 703313P and GRF 
grant 17301115. The second author also thanks the Institute of 
Mathematical Research (IMR), HKU for the financial support of his visit to 
IMR in November, 2017.

\vfill\eject
% ***************************************************************** References

\vfill\eject\end{document}